\documentstyle[PASJadd,epsf]{PASJ95single}
\newcommand{\vol}{}
\newcommand{\no}{}
\newcommand{\lett}{}
\newcommand{\spage}{}
\newcommand{\rdate}{1999 October 26}
\newcommand{\adate}{1999 December 14}

\begin{document}

\title{The Light-cone Effect on the Clustering Statistics \\
in the Cosmological Redshift Space}

\author{Yasushi {\sc Suto},$^{1,2}$ Hiromitsu {\sc Magira}$^1$
\\[12pt]
$^1$ {\it Department of Physics, University of Tokyo, Tokyo
113-0033}\\
{\it E-mail(YS): suto@phys.s.u-tokyo.ac.jp}\\
$^2$ {\it Research Center for the Early Universe, School of Science, 
University of Tokyo, Tokyo 113-0033} \\
and\\
Kazuhiro {\sc Yamamoto} \\
{\it Department of Physics, Hiroshima University, Higashi-Hiroshima 739-8526}}

\abst{
  We present a theoretical formalism to predict the two-point
  clustering statistics (the power spectrum and the two-point
  correlation function), {\it simultaneously} taking account of the
  linear velocity distortion, the nonlinear velocity distortion
  (finger-of-god), the cosmological redshift-space distortion and the
  light-cone effect.  To demonstrate the importance of these effects
  in exploring the clustering of objects at high redshifts, we show
  several model predictions for magnitude-limited surveys of galaxies
  and quasars.  This methodology provides a quantitative tool to
  confront theoretical models against the upcoming precision data on
  clustering in the universe.
}

\kword{ Cosmology: dark matter --- Distance scale --- Large-scale
  structure of universe --- Theory --- Galaxies: distances and
  redshifts }

\maketitle
\thispagestyle{headings}

\section{Introduction}

Observations of clustering of the high-redshift universe suffer from a
variety of {\it contaminations} including the linear redshift-space
(velocity) distortion (Kaiser 1987; Hamilton 1998), the nonlinear
redshift-space (velocity) distortion (Davis \& Peebles 1983; Suto \&
Suginohara 1991; Cole, Fisher, \& Weinberg 1994, 1995), the
cosmological redshift-space (geometrical) distortion (Alcock \&
Paczy\'nski 1979; Ballinger, Peacock, \& Heavens 1996; Matsubara \&
Suto 1996), and the cosmological light-cone effect (Matsubara, Suto,
\& Szapudi 1997; Mataresse et al.  1997; Nakamura, Matsubara, \& Suto
1998; Moscardini et al.  1998; de Laix \& Starkman 1998; Yamamoto \&
Suto 1999; Suto et al. 1999; Nishioka \& Yamamoto 1999).

These effects have been already discussed in the previous papers
quoted above, although separately and not in a unified fashion. In
particular, Yamamoto, Nishioka, \& Suto (1999) derived a formula for
the power spectrum which takes account of the three effects except the
cosmological redshift-space distortion, while Magira, Jing, \& Suto
(2000) extensively discussed the first three effects albeit neglecting
the cosmological light-cone effect.  The former approach requires to
specify the (correct) values of both the density parameter $\Omega_0$
and the cosmological constant $\lambda_0$ (see also Nishioka \&
Yamamoto 2000), and the latter analysis is applicable only for data
over a narrow range of redshift (i.e., $\Delta z/z \ll 1$).

In this Paper, we develop for the first time the relevant formulae for
two-point clustering statistics (the power spectrum and its Fourier
transform, the two-point correlation function) which properly
incorporate all the four effects mentioned above. To demonstrate the
importance of these effects in exploring the clustering of objects at
high redshifts, we present the predictions for future redshift surveys
of galaxies and quasars, assuming spatially-flat cold dark matter
(CDM) cosmology as a representative model.

\section{Formulae for the Two-point Clustering Statistics 
on a Light-cone in the Cosmological Redshift Space}

Our method to predict the two-point clustering statistics of objects
on a light-cone in the cosmological redshift space proceeds as
follows. First we have to specify the model of bias which relates the
density field of objects, $\delta({\bf x},z)$, to that of mass
$\delta_{\rm mass}({\bf x},z)$. Our formulation presented here assumes
the linear bias model (\S 2.1). While the realistic bias model is not
well understood yet, the linear bias model is a fairly good
approximation on large scales independently of the details of the
biasing mechanism (e.g., Coles 1993, Matsubara 1999). Second, we take
into account the redshift-space distortion induced by the peculiar
velocity field (\S 2.2), using both linear theory (Kaiser 1987) and a
non-linear modeling for the finger-of-god effect (Peacock \& Dodd
1996). Third, we correct for the anisotropy due to the geometry of the
universe (\S 2.3). The resulting anisotropic power spectra are
expanded in multipoles and translated to the corresponding moments of
the two-point correlation functions (\S 2.4). Finally those multipoles
defined at a given redshift are integrated over the past light-cone
with the appropriate weight (\S 2.5), which yield the final
expressions for the two-point clustering statistics of objects on a
light-cone in the cosmological redshift space.

\subsection{model of bias}

Since we are mainly interested in the relation between the two-point
statistics on a constant-time hypersurface in the real space and that
on a light-cone hypersurface in the cosmological redshift space, we do
not adopt a realistic, and consequently too complicated, model of
bias. Rather we consider the case of the deterministic, linear and
scale-independent bias:
\begin{eqnarray}
\label{eq:biasdef}
 \delta({\bf x},z)  = 
     b(z)~\delta_{\rm mass}({\bf x},z) ,
\end{eqnarray}
although the real bias might be scale-dependent, nonlinear and
stochastic to some extent (e.g., Taruya, Koyama, \& Soda 1999; Dekel
\& Lahav 1999; Matsubara 1999).  Throughout the Paper, we explicitly
use the subscript, mass, to indicate the quantities related to the
mass density field, and those without the subscript correspond to the
objects satisfying equation (\ref{eq:biasdef}).

\subsection{redshift-space distortion due to the peculiar velocity}

The power spectrum distorted by the peculiar velocity field
(neglecting the cosmological distortion for the moment),
$P^{(S)}(k;z)$, is known to be well approximated by the following
expression (Cole et al. 1995; Peacock \& Dodds 1996):
\begin{eqnarray}
\label{eq:power_in_redshiftspace}
 P^{(S)}(k_\perp,k_\parallel;z) 
  = b^2(z)P^{(R)}_{{\rm\scriptscriptstyle mass}}(k;z) 
\times
\left[1+\beta(z) \left(\frac{k_\parallel}{k}\right)^2 \right]^2
D\left[k_\parallel{\hbox {$\sigma_{\scriptscriptstyle {\rm P}}$}}(z)\right],
\end{eqnarray}
where $k_\perp$ and $k_\parallel$ are the comoving wavenumber
perpendicular and parallel to the line-of-sight of an observer, and
$P^{(R)}_{{\rm\scriptscriptstyle mass}}(k;z)$ is the mass power
spectrum in real space. The second factor in the right-hand-side of
equation (\ref{eq:power_in_redshiftspace}) represents the linear
redshift-space distortion derived by Kaiser (1987) adopting the
distant-observer approximation and the scale-independent deterministic
linear bias.  The parameter $\beta(z)$ is defined by
\begin{eqnarray} 
\label{eq:betaz}
   \beta(z) \equiv {1\over b(z)} \frac{d\ln D_+(z)}{d\ln a} 
\simeq
   {1\over b(z)} \left[\Omega^{0.6}(z) + {\lambda(z) \over 70}
\left(1+ {\Omega(z) \over 2}\right) \right] , 
\end{eqnarray}
where $D_+(z)$ is the gravitational growth rate of the linear density
fluctuations, and $a$ is the cosmic scale factor.  In the above, we
introduce the density parameter, the cosmological constant, and the
Hubble parameter at redshift $z$, which are related to their present
values respectively as
\begin{eqnarray} 
\Omega(z) = \left[{H_0 \over H(z)}\right]^2 \, (1+z)^3 \Omega_0 ,
\quad  \lambda(z) = \left[{H_0 \over H(z)}\right]^2 \, \lambda_0 , \\
 H(z) = H_0\sqrt{\Omega_0 (1 + z)^3 + 
         (1-\Omega_0-\lambda_0) (1 + z)^2 + \lambda_0} .
\end{eqnarray}

For definiteness we adopt an exponential distribution function for the
pairwise peculiar velocity:
\begin{equation}
\label{eq:expveldist}
  f_v(v_{12}) = {1 \over \sqrt{2}{\hbox {$\sigma_{\scriptscriptstyle
         {\rm P}}$}}} \exp\left(-{\sqrt{2}|v_{12}| \over {\hbox
         {$\sigma_{\scriptscriptstyle {\rm P}}$}}} \right) ,
\end{equation}
which leads to the damping function in $k$-space:
\begin{equation}
\label{eq:lorentzdampfactor}
  D[k\mu{\hbox {$\sigma_{\scriptscriptstyle {\rm
  P}}$}}]=\frac{1}{1+(k\mu{\hbox {$\sigma_{\scriptscriptstyle {\rm
  P}}$}})^2/2} ,
\end{equation}
with ${\hbox {$\sigma_{\scriptscriptstyle {\rm P}}$}}$ being the
1-dimensional pair-wise peculiar velocity dispersion.

\subsection{redshift-space distortion due to the geometry of the universe}

Due to a general relativistic effect through the geometry of the
universe, the {\it observable} separations perpendicular and parallel
to the line-of-sight direction, $x_{s{\scriptscriptstyle\perp}} =
(c/H_0)z\delta\theta$ and $x_{s{\scriptscriptstyle\parallel}}=
(c/H_0)\delta z$, are mapped differently to the corresponding comoving
separations in real space $x_{{\scriptscriptstyle\perp}}$ and
$x_{{\scriptscriptstyle\parallel}}$:
\begin{eqnarray}
\label{eq:x2xs}
  x_{s{\scriptscriptstyle\perp}} (z) = x_{\perp} cz/[H_0
  (1+z)d_{\rm\scriptscriptstyle {A}}(z)] \equiv x_{\perp}/c_\perp(z),
\qquad
  x_{s{\scriptscriptstyle\parallel}} (z) = x_{\parallel}
  H(z)/H_0 \equiv x_{\parallel}/c_\parallel(z) ,
\end{eqnarray}
with $d_{\rm\scriptscriptstyle {A}}(z)$ being the angular diameter
distance (Alcock \& Paczy\'nski 1979; Matsubara \& Suto 1996). The
difference between $c_\perp(z)$ and $c_\parallel(z)$ generates the
apparent anisotropy in the clustering statistics which should be
isotropic in the comoving space. Specifically, the power spectrum in
the cosmological redshift space, $P^{({\rm\scriptscriptstyle {CRD}} )}
$, is related to $P^{(S)}$ defined in the {\it comoving} redshift
space (eq.[\ref{eq:power_in_redshiftspace}]) as
\begin{eqnarray}
\label{eq:crdrel}
  P^{({\rm\scriptscriptstyle {CRD}} )}(k_{s\perp},k_{s\parallel};z) 
  =\frac{1}{c_\perp(z)^2c_\parallel(z)} 
\times P^{(S)} \left(\frac{k_{s{\scriptscriptstyle\perp}}}{c_\perp(z)},
\frac{k_{s{\scriptscriptstyle\parallel}}}{c_\parallel(z)};z \right) ,
\end{eqnarray}
where the first factor comes from the Jacobian of the volume element
$dk_{s{\scriptscriptstyle\perp}}^2
dk_{s{\scriptscriptstyle\parallel}}$, and $k_{s\perp}= c_\perp(z)
  k_{\perp}$ and $k_{s\parallel}= c_\parallel(z) k_{\parallel}$ are
  the wavenumber (in the cosmological redshift space) perpendicular
  and parallel to the line-of-sight direction.

Substituting equation (\ref{eq:power_in_redshiftspace}), equation
(\ref{eq:crdrel}) reduces to
\begin{eqnarray}
\label{eq:powercrd}
P^{({\rm\scriptscriptstyle {CRD}} )}(k_s,\mu_k;z)
  &=&\frac{b^2(z)}{c_\perp(z)^2c_\parallel(z)}
P^{(R)}_{\rm\scriptscriptstyle mass} \left(\frac{k_s}{c_\perp(z)}
  \sqrt{1+[{1\over \eta(z)^2}-1]\mu_k^2} ; z \right) \cr &&
  \hspace*{-2cm} \times \left[1+ \left({1\over
  \eta(z)^2}-1\right)\mu_k^2 \right]^{-2} \left[1+ \left({1+\beta(z)
  \over \eta(z)^2}-1\right)\mu_k^2\right]^2 ~ \left[1+
  \frac{k_s^2\mu_k^2{\hbox {$\sigma_{\scriptscriptstyle {\rm
  P}}$}}^2}{2c^2_\parallel(z)} \right]^{-1},
\end{eqnarray}
where we introduce
\begin{eqnarray}
\label{eq:k2ks}
k_s \equiv \sqrt{ k_{s\perp}^2 + k_{s\parallel}^2}, \quad
  \mu_k \equiv k_{s\parallel}/k_s, \quad
  \eta \equiv c_\parallel/c_\perp,
\end{eqnarray}
(Ballinger, Peacock, \& Heavens 1996; Magira, Jing, \& Suto 2000).

The two-point correlation function in the cosmological redshift space,
$\xi^{({\rm\scriptscriptstyle {CRD}} )}(x_{s\perp},x_{s\parallel};z)$,
is computed using equation (\ref{eq:powercrd}) as
\begin{eqnarray}
\xi^{({\rm\scriptscriptstyle {CRD}} )}({\bf x_s};z)
&=& {1 \over (2\pi)^3} 
\int P^{({\rm\scriptscriptstyle {CRD}} )}({\bf k_s};z) 
\exp(-i{\bf k_s}\cdot{\bf x_s}) d^3k_s \cr
&=& {1 \over (2\pi)^3} \int 
  P^{(S)}({\bf k};z) \exp(-i{\bf k}\cdot{\bf x}) d^3k \cr
&=& \xi^{(S)}(c_\perp x_{s\perp},c_\parallel x_{s\parallel};z) ,
\end{eqnarray}
where $\xi^{(S)}(x_{\perp}, x_{\parallel};z)$ is the redshift-space
correlation function defined through equation
(\ref{eq:power_in_redshiftspace}).

\subsection{multipole expansion}

Following Hamilton (1992), we decompose the power spectrum into
harmonics:
\begin{eqnarray}
\label{eq:pkmoment}
P(k,\mu_k;z) = \sum_{l: even} P_l(k) L_l(\mu_k), \quad 
P_l(k;z) \equiv 
\frac{2l+1}{2}\int^1_{-1}d\mu_k P(k,\mu_k;z) L_l(\mu_k) ,
\end{eqnarray}
where $\mu_k$ is the direction cosine between the wavevector and the
line-of-sight, and $L_l(\mu_k)$ are the $l$-th order Legendre
polynomials. Similarly, the two-point correlation function is
decomposed as
\begin{eqnarray}
\label{eq:ximoment}
\xi(x,\mu_x;z) = \sum_{l: even} \xi_l(x) L_l(\mu_x), \quad 
\xi_l(x;z) \equiv 
\frac{2l+1}{2}\int^1_{-1}d\mu_x \xi(x,\mu_x;z) L_l(\mu_x) ,
\end{eqnarray}
using the direction cosine $\mu_x$ between the separation vector and
the line-of-sight.

The above multipole moments satisfy the following relations:
\begin{eqnarray}
\label{eq:pk2xi}
P_l(k;z) = 4\pi i^l \int_0^\infty \xi_l(x;z) j_l(kx) x^2dx, \quad 
\xi_l(x;z) = {1 \over 2\pi^2 i^l} \int_0^\infty P_l(k;z) j_l(kx) k^2dk,
\end{eqnarray}
with $j_l(kx)$ being the spherical Bessel functions. Substituting
$P^{({\rm\scriptscriptstyle {CRD}} )}(k_s,\mu_k;z)$ and
$\xi^{({\rm\scriptscriptstyle {CRD}} )}({\bf x_s};z)$ in equations
(\ref{eq:pkmoment}) to (\ref{eq:pk2xi}), one can compute the moments,
$P^{({\rm\scriptscriptstyle {CRD}} )}_l(k_s;z)$ and
$\xi^{({\rm\scriptscriptstyle {CRD}} )}_l(x_s;z)$ at a given $z$.
Actually the above relations are essential in predicting the two-point
correlation function in the cosmological redshift space with the
nonlinear effects.

\subsection{the light-cone effect}

Finally we incorporate the light-cone effect following the expression
by Yamamoto \& Suto (1999).  Since $P_l^{({\rm\scriptscriptstyle
{CRD}} )}(k_s;z)$ and $\xi_l^{({\rm\scriptscriptstyle {CRD}} )}
(k_s;z)$ are defined in redshift space, the proper weight should be
\begin{eqnarray}
\label{eq:weight}
d^3s^{({\rm\scriptscriptstyle {CRD}} )}
~[\phi(z)n_0^{{\rm\scriptscriptstyle {CRD}} }(z)]^2
=d^3x{[\phi(z)n_0^{{\rm\scriptscriptstyle {com}} }(z)]^2
~c_\perp(z)^2c_\parallel(z)},
\end{eqnarray}
where $n_0^{{\rm\scriptscriptstyle {CRD}} }(z)$ and
$n_0^{{\rm\scriptscriptstyle {com}} }(z)$ denote number densities of
the objects in cosmological redshift space and comoving space,
respectively, and $\phi(z)$ is the selection function determined by
the observational target selection and the luminosity function of the
objects. Then the final expressions reduce to
\begin{eqnarray}
\label{eq:lccrdpkmom}
    P^{({\rm\scriptscriptstyle {LC}} ,{\rm\scriptscriptstyle {CRD}} )}_l(k_s) 
&=& {
   {\displaystyle 
     \int_{z_{\rm max}}^{z_{\rm min}} dz 
     {dV_{\rm c} \over dz} ~[\phi(z)n_0^{{\rm\scriptscriptstyle {com}} }(z)]^2
    {c_\perp(z)^2c_\parallel(z)} P_l^{({\rm\scriptscriptstyle {CRD}} )}(k_s;z)
    }
\over
    {\displaystyle
     \int_{z_{\rm max}}^{z_{\rm min}} dz 
     {dV_{\rm c} \over dz}  ~[\phi(z)n_0^{{\rm\scriptscriptstyle {com}} }(z)]^2
          {c_\perp(z)^2c_\parallel(z)}
     }
} , \\
\label{eq:lccrdximom}
    \xi^{({\rm\scriptscriptstyle {LC}} ,{\rm\scriptscriptstyle {CRD}} )}_l(x_s) 
&=& {
   {\displaystyle 
     \int_{z_{\rm max}}^{z_{\rm min}} dz 
     {dV_{\rm c} \over dz} ~[\phi(z)n_0^{{\rm\scriptscriptstyle {com}} }(z)]^2
          {c_\perp(z)^2c_\parallel(z)}
    \xi^{{\rm\scriptscriptstyle {CRD}} }_l(x_s;z)
    }
\over
    {\displaystyle
     \int_{z_{\rm max}}^{z_{\rm min}} dz 
     {dV_{\rm c} \over dz}  ~[\phi(z)n_0^{{\rm\scriptscriptstyle {com}} }(z)]^2
            {c_\perp(z)^2c_\parallel(z)}
     }
} ,
\end{eqnarray}
where $z_{\rm max}$ and $z_{\rm min}$ denote the redshift range of the
survey, $dV_{\rm c}/dz = d_{\rm\scriptscriptstyle {C}} ^2(z)/H(z)$ is
the comoving volume element per unit solid angle.

Note that $k_s$ and $x_s$ defined in $P_l^{({\rm\scriptscriptstyle
    {CRD}} )}(k_s;z)$ and $\xi^{{\rm\scriptscriptstyle {CRD}} }_
l(x_s;z)$ are related to their comoving counterparts at $z$ through
equations (\ref{eq:k2ks}) and (\ref{eq:x2xs}) while those in
$P^{({\rm\scriptscriptstyle {LC}} ,{\rm\scriptscriptstyle {CRD}} )}_
l(k_s)$ and $\xi^{({\rm\scriptscriptstyle {LC}} ,
  {\rm\scriptscriptstyle {CRD}} )}_l(x_s)$ are not specifically
related to any comoving wavenumber and separation. Rather they
correspond to the quantities averaged over the range of $z$ satisfying
the observable conditions of $x_s=(c/H_0)\sqrt{\delta z^2 +
  z^2\delta\theta^2}$ and $k_s=2\pi/x_s$.

\section{Predictions for Galaxy and QSO Samples in CDM Models}

As in Yamamoto, Nishioka, \& Suto (1999), we consider SCDM (standard
cold dark matter) and LCDM (Lambda cold dark matter) models, which
have $(\Omega_0, \lambda_0, h, \sigma_8)$ $= (1.0, 0.0, 0.5, 0.6)$ and
$(0.3, 0.7, 0.7, 1.0)$, respectively. These sets of cosmological
parameters are chosen so as to reproduce the observed cluster
abundance (Kitayama \& Suto 1997).  We use the fitting formulae of
Peacock \& Dodds (1996) and Mo, Jing, \& B\"{o}rner (1997) for the
nonlinear power spectrum $P^{(R)}_{{\rm\scriptscriptstyle {mass}} }
(k;z)$ and the peculiar velocity dispersions ${\hbox
  {$\sigma_{\scriptscriptstyle {\rm P}}$}}$, respectively (see also
Suto et al. 1999; Magira et al. 2000). Also we take into account the
selection functions relevant to the upcoming SDSS spectroscopic
samples of galaxies and quasars adopting the B-band limiting
magnitudes of 19 and 20, respectively.  Unlike Yamamoto \& Suto (1999)
and Yamamoto, Nishioka, \& Suto (1999), we properly transform the QSO
luminosity function originally fitted assuming the Einstein -- de
Sitter universe (Wallington \& Narayan 1993; Nakamura \& Suto 1997) to
that in a given cosmological model.

\begin{figure}[thbp]
\begin{center}
  \leavevmode\epsfxsize=9cm \epsfbox{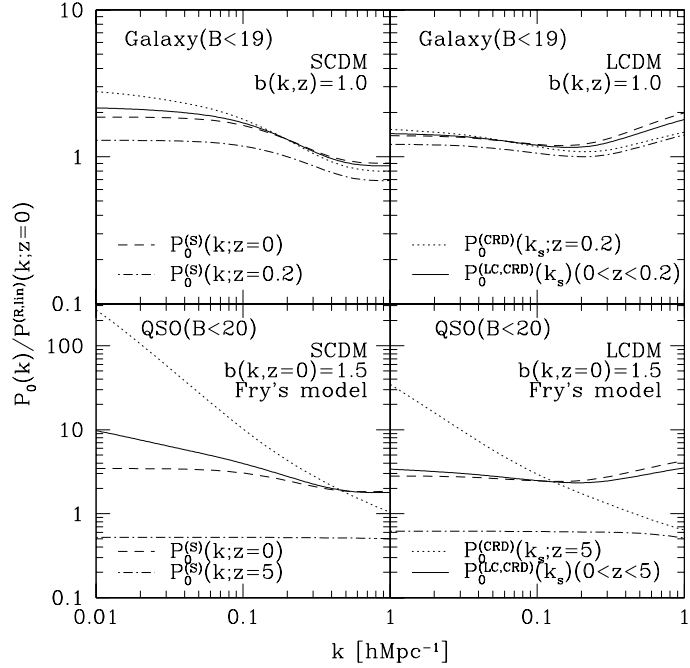}
\end{center}
\caption{ The light-cone and cosmological redshift-space distortion
  effects on angle-averaged power spectra.  Upper and lower panels
  correspond to magnitude-limited samples of galaxies ($B<19$ in
  $0<z<{z_{\rm max}}=0.2$; no bias model) and QSOs ($B<20$ in
  $0<z<{z_{\rm max}}=5$; Fry's linear bias model), respectively.  {\it
    Dashed}: $P^{\rm (S)}_0(k;z=0)/P^{\rm (R,lin)}(k;z=0)$, {\it
    dash-dotted}: $P^{\rm (S)}_0(k;z={z_{\rm max}})$, {\it dotted}:
  $P^{({\rm\scriptscriptstyle {CRD}} )}_0(k_s;z={z_{\rm max}})$, {\it
    solid}: $P^{({\rm\scriptscriptstyle {LC}} ,{\rm\scriptscriptstyle
      {CRD}} )}_0(k_s)$.
\label{fig:pk_lccrd}
}
\end{figure}

Figure 1 compares the predictions for the angle-averaged (monopole)
power spectra under various approximations.  Upper and lower panels
adopt the selection functions appropriate for galaxies in $0<z<{z_{\rm
max}}=0.2$ and QSOs in $0<z<{z_{\rm max}}=5$, respectively.  Left and
right panels present the results in SCDM and LCDM models. For
simplicity we adopt a scale-independent linear bias model of Fry
(1996):
\begin{equation} 
  b(z)= 1 +{1\over D_+(z)} [b(k,z=0)-1],
\label{FryM}
\end{equation}
with $b(k,z=0)=1$ and $1.5$ for galaxies and quasars, respectively.
We present the results normalized by the real-space power spectrum in
linear theory (Bardeen et al. 1986); $P^{\rm (S)}_0(k;z=0)$, $P^{\rm
(S)}_0(k;z={z_{\rm max}})$, $P^{({\rm\scriptscriptstyle {CRD}} )}_
0(k_s;z={z_{\rm max}})$, and $P^{({\rm\scriptscriptstyle {LC}} ,
{\rm\scriptscriptstyle {CRD}} )}_0(k_s)$ are computed using the
nonlinear power spectrum of Peacock \& Dodds (1996), and plotted in
dashed, dash-dotted, dotted and solid lines, respectively.

Consider the results for the galaxy sample (upper panels) first. On
linear scales ($k < 0.1h$Mpc$^{-1}$), $P^{\rm (S)}_0(k;z=0)$ plotted
in dashed lines is enhanced relative to that in real space mainly due
to the linear redshift-space distortion (The Kaiser factor in
eq.[\ref{eq:power_in_redshiftspace}]). For nonlinear scales, the
nonlinear gravitational evolution increases the power spectrum in real
space while the finger-of-god effect suppresses that in redshift
space. Thus the net result is sensitive to the shape and the amplitude
of the fluctuation spectrum itself; in the LCDM model that we adopted,
the nonlinear gravitational growth in real space is stronger than the
suppression due to the finger-of-god effect. Thus $P^{\rm
  (S)}_0(k;z=0)$ becomes larger than its real-space counterpart in
linear theory. In the SCDM mode, however, this is opposite and $P^{\rm
  (S)}_0(k;z=0)$ becomes smaller.

The power spectra at $z=0.2$ (dash-dotted lines) are smaller than
those at $z=0$ by the corresponding growth factor of the fluctuations,
and one might expect that the amplitude of the power spectra on the
light-cone (solid lines) would be in-between the two.  While this is
correct if we use the comoving wavenumber, the actual observation on
the light-cone in the cosmological redshift space should be expressed
in terms of $k_s$ (eq.[(\ref{eq:k2ks}]).  If we plot the power spectra
at $z=0.2$ taking into account the geometrical distortion,
$P^{({\rm\scriptscriptstyle {CRD}} )}_0(k_s;z=0.2)$ in dotted lines
becomes significantly larger than $P^{\rm (S)}_0(k;z=0.2)$. Therefore
$P^{({\rm\scriptscriptstyle {LC}} , {\rm\scriptscriptstyle {CRD}} )}_
0(k_s)$ should take a value between those of
$P^{({\rm\scriptscriptstyle {CRD}} )}_0(k_s;z=0) = P^{\rm
  (S)}_0(k;z=0)$ and $P^{({\rm\scriptscriptstyle {CRD}} )}_
0(k_s;z=0.2)$. This explains the qualitatively features shown in upper
panels of figure 1.  As a result, both the cosmological redshift-space
distortion and the light-cone effect substantially change the
predicted shape and amplitude of the power spectra, even for the
galaxy sample (Nakamura, Matsubara, \& Suto 1998).

The results for the QSO sample can be basically understood in a
similar manner except that the evolution of bias makes significant
difference since the sample extends to much higher redshifts.
Although it is true that the results are sensitive to the model of
bias, we believe that those on linear scales are fairly insensitive to
the complexities of the bias; it has been shown that the bias on such
scales can be well approximated by the linear and scale-independent
model (Coles 1993; Matsubara 1999; Magira, Taruya, Jing \& Suto 2000).

\begin{figure}[thbp]
\begin{center}
   \leavevmode\epsfxsize=9cm \epsfbox{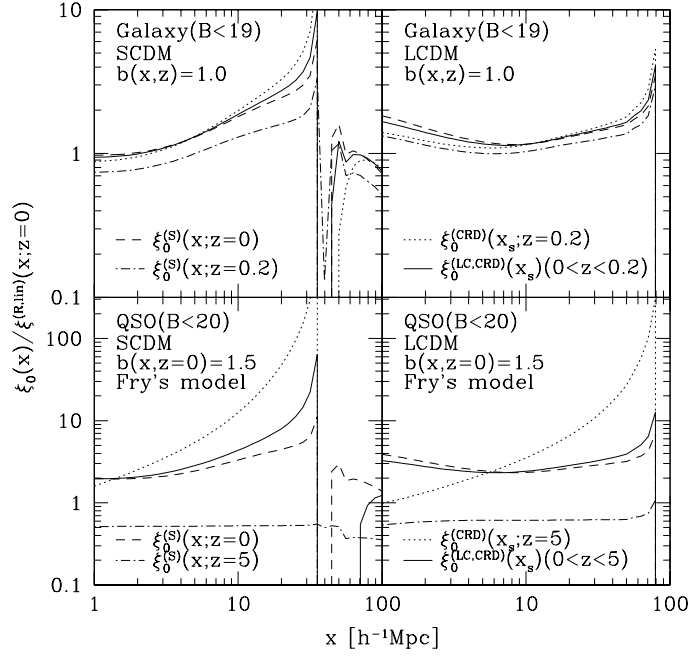}
\end{center}
\caption{The same as Fig.\ref{fig:pk_lccrd} on angle-averaged
  two-point correlation functions.
\label{fig:xi_lccrd}
}
\end{figure}

Figure 2 shows the results for the angle-averaged (monopole) two-point
correlation functions, exactly corresponding to those in figure 1. The
results in this figure can be also understood by the analogy of those
presented in figure 1 at $k \sim 2\pi/x$. Unlike the power spectra,
however, two-point correlation functions are not positive definite.
The funny features in figure 2 on scales larger than $30 h^{-1}$Mpc
($100 h^{-1}$Mpc) in SCDM (LCDM) originate from the fact that
$\xi^{\rm (R,lin)}(x,z=0)$ becomes negative there.

In the linear and deterministic bias model that we adopted here, the
power spectra of objects at $z$ in redshift space (without the
geometrical effect) are proportional to the mass power spectrum in
real space, $P^{(R)}_{\rm\scriptscriptstyle mass}(k;z)$.  Hamilton
(1992) showed that in linear theory of redshift-space distortion
(Kaiser 1987), the quadrupole to monopole ratio provides a measure of
estimating $\beta$ through the following relation:
\begin{eqnarray}
\label{eq:p2p0lin}
{P^{\rm S,lin}_2(k;z) \over P^{\rm S,lin}_0(k;z)}
= \frac{\displaystyle {4\over 3}\beta(z) + {4\over 7}\beta^2(z)}
       {\displaystyle 1+{2\over 3}\beta(z) + {1\over 5}\beta^2(z)  }
= \frac{\displaystyle \xi^{\rm S,lin}_2(x;z)}
     {\displaystyle \xi^{\rm S,lin}_0(x;z) 
            -3 \int_0^1  \xi^{\rm S,lin}_0(xw;z) w^2dw} .
\end{eqnarray}
Taking account of the nonlinear redshift-space distortion, however,
the ratio becomes scale-dependent; our present model
(\ref{eq:power_in_redshiftspace}) with equation
(\ref{eq:lorentzdampfactor}) yields
\begin{eqnarray}
\label{eq:p2p0nl}
{P^{\rm S}_2(k;z) \over P^{\rm S}_0(k;z)}
&=& \frac{\displaystyle 
{5\over2} \left[B(\kappa)-A(\kappa)\right]
  +\beta(z)\left[3C(\kappa)- {5\over3}B(\kappa)\right]
 +\beta^2(z)\left[{3\over2\kappa^2}(1-C(\kappa))-{1\over2}C(\kappa)\right] 
} 
{\displaystyle 
A(\kappa)+{2\over3}\beta(z) B(\kappa)+{1\over 5}\beta^2(z) C(\kappa)
}  ,  \\
 A(\kappa) &=& {{\arctan}(\kappa) \over \kappa},
\quad
  B(\kappa) = {3\over\kappa^2}
    \left[1-{{\arctan}(\kappa) \over \kappa}\right] ,
\quad
  C(\kappa) = {5\over3\kappa^2}\left[1-{3\over\kappa^2} +
{3{\arctan}(\kappa) \over \kappa^3} \right] ,
\end{eqnarray}
with $\kappa(z)=k{\hbox {$\sigma_{\scriptscriptstyle {\rm
        P}}$}}(z)/\sqrt{2}H_0$ (Magira, Jing, \& Suto 2000; Yamamoto,
Nishioka, \& Suto 1999; see also Cole, Fisher, \& Weinberg 1995).  We
note here that $P^{\rm S}_2(k;z)$ is not any more positive definite
and in fact changes sign at cosmologically interesting scales (see
figure 3 below).

The above ratio of monopole and quadrupole turns out to be independent
of the underlying mass power spectrum, which is regarded as one of the
advantages of estimating the $\beta$-parameter using the
redshift-space distortion.  The cosmological redshift-space distortion
(eqs.[\ref{eq:powercrd}] and [\ref{eq:pkmoment}]) and the light-cone
effect (eqs.[\ref{eq:lccrdpkmom}] and [\ref{eq:lccrdximom}]), however,
further complicate the behavior of the final predictions for the
two-point statistics, which are plotted in figure 3.  In fact, even
the shape of $P^{({\rm\scriptscriptstyle {LC}} ,{\rm\scriptscriptstyle
    {CRD}} )}_ 0(k_s)$ and $P^{({\rm\scriptscriptstyle {LC}} ,
  {\rm\scriptscriptstyle {CRD}} )}_2(k_s)$ for QSOs is sensitive to
the bias model and the cosmological parameters.

For the galaxy sample, $P^{({\rm\scriptscriptstyle {LC}} ,
  {\rm\scriptscriptstyle {CRD}} )}_ 2(k_s)/P^{({\rm\scriptscriptstyle
    {LC}} , {\rm\scriptscriptstyle {CRD}} )}_0(k_s)$ in large scales
approaches 1.06 (0.65) for SCDM (LCDM), which should be compared with
$50/49$ (0.57) in linear theory (eq.[\ref{eq:p2p0lin}]). This implies
that such estimated values of $\beta$ are 1.05 (0.57), in contrast to
the true values of 1.0 (0.49). Thus even for the shallow samples of
galaxies, the present effects systematically changes the estimate of
$\beta$ by $(5\sim20)$\% level (Nakamura, Matsubara, \& Suto 1998).

\begin{figure}[thbp]
\begin{center}
  \leavevmode\epsfxsize=9cm \epsfbox{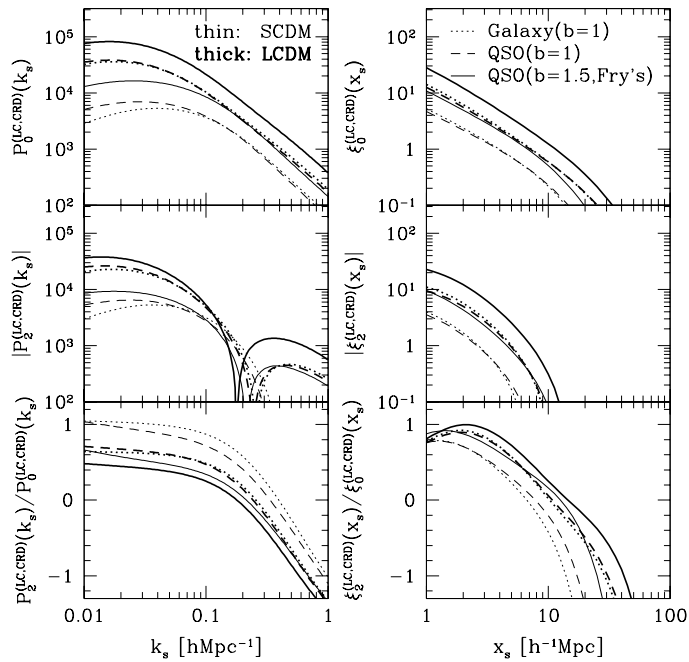}
\end{center}
\caption{Predictions for the two-point statistics on a light-cone
  in the cosmological redshift space.  The results in SCDM and LCDM
  models are plotted in thin and thick lines, respectively.
{\it Left:} monopole, quadrupole and
  quadrupole-to-monopole ratio of the power spectrum (from top to
  bottom).  {\it Right:} monopole, quadrupole and
  quadrupole-to-monopole ratio of the two-point correlation functions
  (from top to bottom).
\label{fig:mono2quad_lccrd}
}
\end{figure}

\section{Conclusions and Discussion}

We have developed for the first time a theoretical formalism to
predict the two-point clustering statistics on a light-cone in the
cosmological redshift space, $P^{({\rm\scriptscriptstyle {LC}} ,
{\rm\scriptscriptstyle {CRD}} )}(k_s)$ and
$\xi^{({\rm\scriptscriptstyle {LC}} ,{\rm\scriptscriptstyle {CRD}} )}_
0(x_s)$, which were only discussed separately before (Yamamoto,
Nishioka, \& Suto 1999; Magira, Jing, \& Suto 2000). This methodology
fully takes into account all the {\it well-defined} physical, i.e.,
gravitational and dynamical effects, and
enables one to quantitatively confront the future precision data on
the clustering statistics against the cosmological model predictions
{\it modulo} the uncertainty of the bias.  

In fact, since the resulting predictions are sensitive to the bias,
which is unlikely to quantitatively be specified by theory, the
present methodology will find two completely different applications.
For relatively shallower catalogues like galaxy samples, the evolution
of bias is not supposed to be so strong. Thus one may estimate the
cosmological parameters from the observed degree of the redshift
distortion as has been conducted conventionally. Most importantly we
can correct for the systematics due to the light-cone and geometrical
distortion effects which affect the estimate of the parameters by
$\sim 10$\% (\S 3).  This is generally the case on linear and
quasi-nonlinear scales where it has been shown that the bias on such
scales can be well approximated by the linear and scale-independent
model (Coles 1993; Matsubara 1999; Magira, Taruya, Jing \& Suto 2000)
as we have assumed in this paper.

Alternatively, for deeper catalogues like high-redshift quasar
samples, one can extract information on the nonlinearity,
scale-dependence and stochasticity of the object-dependent bias
(Taruya, Koyama, \& Soda 1999; Dekel \& Lahav 1999; Blanton et al.
1999) only by correcting the observed data on the basis of our
formulae. In this case one should adopt a set of cosmological
parameters a priori, but they will be provided both from the
low-redshift analysis described above and from the precision data of
the cosmic microwave background and supernovae Ia.

In a sense, the former approach uses the light-cone and geometrical
distortion effects as real cosmological signals, while the latter
regards them as inevitable, but physically removable, noises. In both
cases, the present methodology is essential in properly interpreting
the observations of the universe at high redshifts.

\vspace{1pc}\par
We thank Shin Sasaki for suggesting the proper scaling of the QSO
luminosity function to an arbitrary cosmological model.  This research
was supported in part by the Grants-in-Aid by the Ministry of
Education, Science, Sports and Culture of Japan to RESCEU (07CE2002)
and to K.Y.  (11640280), and by the Inamori Foundation.


\section*{References}

\re
  Alcock C., Paczy\'nski B.\ 1979, Nature 281, 358
\re
  Ballinger W.E., Peacock J.A., Heavens A.F.\ 1996, MNRAS 282, 877
\re
  Bardeen J.M., Bond J.R., Kaiser N., Szalay A.S.\ 1986, ApJ 304, 15
\re
  Blanton M., Cen R.Y., Ostriker J.P., Strauss M.A.\ 1999, ApJ 522, 590
\re
  Coles P.\ 1993, MNRAS 262, 1065
\re
  Cole S., Fisher K.B., Weinberg D.H.\ 1994, MNRAS 267, 785
\re
  Cole S., Fisher K.B., Weinberg D.H.\ 1995, MNRAS 275, 515
\re
  Davis M., Peebles P.J.E.\ 1983, ApJ 267, 465
\re
  Dekel A., Lahav O.\ 1999, ApJ 520, 24
\re
  de Laix A.A., Starkman G.D.\ 1998, MNRAS 299, 977
\re
  Fry J.N.\ 1996, ApJ 461, L65
\re
  Hamilton A.J.S.\ 1992, ApJ 385, L5
\re
  Hamilton A.J.S.\ 1998, The Evolving Universe. Selected Topics
  on Large-Scale Structure and on the Properties of Galaxies
  (Kluwer, Dordrecht) p185
\re
  Kaiser N.\ 1987, MNRAS 227, 1
\re
  Kitayama T., Suto Y.\ 1997, ApJ 490, 557
\re
  Magira H., Jing Y.P., Suto Y.\ 2000, ApJ 528, January 1 issue, 
   in press (astro-ph/9907438)
\re
  Magira H., Taruya, A., Jing Y.P., Suto Y.\ 2000, in preparation
\re
  Matarrese S., Coles P., Lucchin F., Moscardini L.\ 1997, MNRAS 286, 115
\re
  Matsubara T.\ 1999, ApJ 525, 543
\re
  Matsubara T., Suto Y.\ 1996, ApJ 470, L1
\re
  Matsubara T., Suto Y., Szapudi I.\ 1997, ApJ 491, L1
\re
  Mo H.J., Jing Y.P., B\"orner G.\ 1997, MNRAS 286, 979 
\re
  Moscardini L., Coles P., Lucchin F., Matarrese S.\ 1998, MNRAS 299, 95
\re
  Nakamura T.T., Matsubara T., Suto Y.\ 1998, ApJ 494, 13
\re
  Nakamura T.T., Suto Y. 1997, Prog.\ Theor.\ Phys. 97, 49
\re
  Nishioka H., Yamamoto K.\ 1999, ApJ 520, 426
\re
  Nishioka H., Yamamoto K.\ 2000, ApJS, in press
\re
  Peacock J.A., Dodds S.J.\ 1996, MNRAS 280, L19
\re
  Suto Y., Magira H., Jing Y.P., Matsubara T.,  Yamamoto K.\ 1999,
  Prog.\ Theor.\ Phys.\ Suppl. 133, 183
\re
  Suto Y., Suginohara T.\ 1991, ApJ 370, L15
\re
  Taruya A., Koyama K., Soda J.\ 1999, ApJ 510, 541
\re
  Wallington S., Narayan R.\ 1993, ApJ 403, 517
\re
  Yamamoto K., Nishioka H., Suto Y.\ 1999, ApJ 527, December 20
issue, in press (astro-ph/9908006)
\re
  Yamamoto K., Suto Y.\ 1999, ApJ 517, 1
\re


\label{last}
\end{document}